\documentclass[aps,prl,reprint,twocolumn,showpacs,epsf]{revtex4}
\usepackage{graphicx}
\usepackage{psfrag}
\usepackage{color}

\def\a{\alpha}

\def\ve{\varepsilon}
\def\g{\gamma}

\begin{document}
\bibliographystyle{apsrev}

\hspace{5cm} {\it Phys. Rev. Lett.} {\bf 95}, 170603, (2005)
\title{First Passage and Cooperativity of Queuing Kinetics}



\author{Maria R. D'Orsogna and Tom Chou}

\affiliation{Dept. of Mathematics, UCLA, Los Angeles, CA 90095}
\affiliation{Dept. of Biomathematics, UCLA, Los Angeles, CA 90095}


\date{\today}

\begin{abstract}
\noindent
We model the kinetics of ligand-receptor systems, where multiple
ligands may bind and unbind to the receptor, either randomly or in a
specific order.  Equilibrium occupation and first occurrence of
complete filling of the receptor are determined and compared.  At
equilibrium, receptors that bind ligands sequentially are more likely
to be saturated than those that bind in random order.  Surprisingly
however, for low cooperativity, the random process first reaches full
occupancy faster than the sequential one.  This is true {\it except}
near a critical binding energy where a 'kinetic trap' arises and the
random process dramatically slows down when the number of binding
sites $N\geq 8$.  These results demonstrate the subtle interplay
between cooperativity and sequentiality for a wide class of kinetic
phenomena, including chemical binding, nucleation, and assembly line
strategies.
\end{abstract}

\maketitle

Cooperativity plays a key role in determining the equilibrium
properties of queuing systems such as ligand-receptor binding,
nucleation, melting of $\alpha$-helices, coupled chemical reactions,
and assembly lines.  Biophysical examples include $O_{2}$ or $CO$
binding to hemoglobin and myoglobin \cite{HEMO1,HEMO2,HANGGI} and
binding during cell signaling and morphogenesis \cite{MORPH1,MORPH2}.
For a local bulk ligand concentration, the associated receptors will
typically have a fraction of sites filled.
The kinetics of queuing in these processes can exhibit diverse and
rich behavior.  A receptor may need to have a critical number of bound
ligands before it can signal the next biochemical step.  Thus, it is
important to know not only the equilibrium ligand occupancy, but also
the mean time to first reach this critical occupancy, as a function of
local ligand concentration and binding strength. Similarly, in
nucleation processes such as $\alpha$-helix formation or melting,
local helix turns can form randomly or sequentially. The first time a
complete helix forms (or melts) will be an important ingredient in
protein folding models \cite{HELIX}. First passage times also define
extinction and fixation in birth-death processes
\cite{DOERING,REDNER}.  Equilibrium distributions and first passage
times also arise in applications of queuing, where, for example,
average computer loads and the first time that demand exceeds capacity
should be distinguished \cite{QUEUE1}.

\begin{figure}
\begin{center}
\includegraphics[height=1.22in]{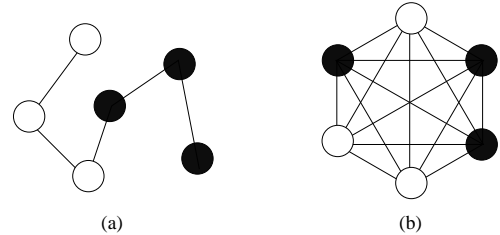}
\end{center}
\caption{Two limiting models of multiple ligand binding. In both
cases, $N=6$, and $n=3$. (a) Sequential ligand binding/unbinding. At
any given time, only one specific site can accept ligand adsorption
and only one ligand can desorb. Upon defining the kinetics, this case
will be denoted $\a=0$. (b) Random ligand binding/unbinding.  An
additional ligand can bind to any one of the open sites.  Similarly,
any one of the bound ligands can spontaneously desorb. This case will
be denoted $\a=1$.}
\label{Fig1}
\end{figure}

In this Letter, we formulate and use a kinetic chain model to
highlight subtleties of ligand adsorption and desorption, queuing, and
cooperativity.  Our model is presented in the language of ligand
binding to a single receptor with $N$ active sites of which $0 \leq n
\leq N$ are occupied by ligands at any given time.  The {\it order} of
the binding can be imposed in two limiting ways.  As shown in
Fig.\,\ref{Fig1}, the addition of each successive ligand can influence
one other specific site and allow the next ligand to bind to, or
unbind from, that site only (a case we will denote by the index
$\a=0$). Alternatively, the allosteric effect (from {\it e.g.} a large
scale conformational change) can be spread equally to all remaining
sites.  The next ligand can bind to any one of these remaining open
sites (a case we will denote by the index $\a=1$).  Here, all
bound ligand molecules are equally likely to spontaneously desorb. We
do not consider mixed processes in which the binding order is
sequential and the unbinding random, or {\it vice versa}.  Thus, in
our model, ligand binding occurs in a totally sequential manner as in
Fig.\,\ref{Fig1}a, or randomly as shown in Fig.\,\ref{Fig1}b.  We show
how the binding order plays a crucial role in determining both
equilibrium and kinetic properties: sequential ordering generally
implies higher occupancy, while random ordering is associated with a
shorter mean first passage to saturation. We also discover an
intriguing regime of kinetic slowdown where different queuing rules
result in dramatically different behaviors.

We define $P(n,t)$ as the probability that the receptor has $n$ bound
ligands at time $t$, given that it had $n_{0}$ at time $t_{0}$.  The
evolution of the occupancy state $n$ can be mapped onto a
one-dimensional random walk with a master equation given by
$\partial_{t}P(n,t) = q_{n+1} P(n+1,t) - (q_n +k_n) P(n,t) + k_{n-1}
P(n-1,t)$.  Here, $k_n$ and $q_n$ represent the ligand adsorption and
desorption rates, respectively, when there are already $n$ ligands
bound to the receptor.  The probability density current from state
$n+1$ to state $n$ is $J(n,t) = q_{n+1} P(n+1,t) - k_n P(n,t)$.  The
master equation, $\partial_{t}P(n,t) = J(n,t) -J(n-1,t)$, 
is solved under the constraint $\sum_{i=0}^{N}P_{i} = 1$.

Different functional forms of $k_{n}$ and $q_{n}$ distinguish the
sequential model from the random one.  The number of ways for a ligand
to bind an $n$ state receptor is proportional to the number of
accessible binding sites.  For sequential kinetics, at any given time,
only one site is accessible for binding the next ligand, whereas for
random binding, all $N-n$ empty sites are accessible. The number of
accessible sites can thus be succinctly expressed as $(N-n)^{\a}$,
where the index $\a = 0$ corresponds to the sequential model, and
$\a=1$ corresponds to the random one.  Similarly, the number ways to
detach a particle from an occupation state $n+1$ is $(n+1)^\a$.  If
conformational changes of the receptor molecule reach local
thermodynamic equilibrium within the time scales required for ligand
binding and unbinding, the rates obey detailed balance:

\vspace{-5mm}

\begin{equation}
{k_{n} \over q_{n+1}} = 
z \left({N-n \over n+1}\right)^{\alpha}\exp\left[-\Delta G_{n} \right].
\label{RATES}
\end{equation}
The proportionality constant $z\approx v [L]$ is dimensionless and
depends on the bulk ligand concentration $[L]$ and on the capture
volume per binding site $v$.  Typical values of $\left[ L \right] =
1\mu M$ and $ v = 1\,nm ^{3}$ give $z \approx 10^{-6}$.  The quantity
$ - \Delta G_{n} \equiv G_{n} - G_{n+1}$ is the free energy change, in
units of thermal energy $k_B T$, upon detachment of one ligand from
state $n+1$ to state $n$.  Equivalently, $\Delta G_{n}$ can be
interpreted as the free energy change due to ligand addition from
state $n$ to state $n+1$.  Cooperativity can be defined as the
additional proclivity for the $(n+1)^{st}$ ligand to bind as $n$
increases. A simple model for the free energy of binding is $\Delta
G_{n} = -\varepsilon_{0} (n+1)^{\gamma}$ where $\varepsilon_0$ is the
nonnegative ligand binding energy in units of $k_{B}T$.  The
nonnegative parameter $\gamma$ controls the cooperativity of
successive binding.  For $\gamma = 0 $, $\Delta G_n = -\varepsilon_0$
is independent of $n$ and the binding is noncooperative.  If $\gamma >
0$, $\Delta G_n$ decreases with $n$, hence a positive cooperative
effect. Large values of $\gamma$ represent strong cooperativity.  The
equilibrium probability distribution $P_{eq}(n)$ is derived by
imposing $J(n,t)=0$ and by using reflecting boundary conditions for
the empty and full states, $k_{-1}, q_{0} =0$ and $q_{N+1}, k_{N}=0$,
respectively \cite{VANKAMPEN}.  Upon using Eq.\,\ref{RATES},

\vspace{-2mm}
\begin{equation}
\displaystyle P_{eq}(n) = \frac{
\displaystyle{ z^{n} {{N} \choose {n}}^{\a} \prod_{s=1}^{n} \exp
\left[{\epsilon_0 \, s^\gamma}\right] \,(1-\delta_{n,0}) 
+ \delta_{n,0} }}
{\displaystyle{1+ \sum_{m=1}^{N}z^{m} {{N} \choose {m}}^{\a}
\prod_{s=1}^{m} \exp \left[{\epsilon_0 \, s^\gamma}\right]}}.
\label{PEQ}
\end{equation}

\begin{figure}[t]
\begin{center}
\includegraphics[height=3.3in]{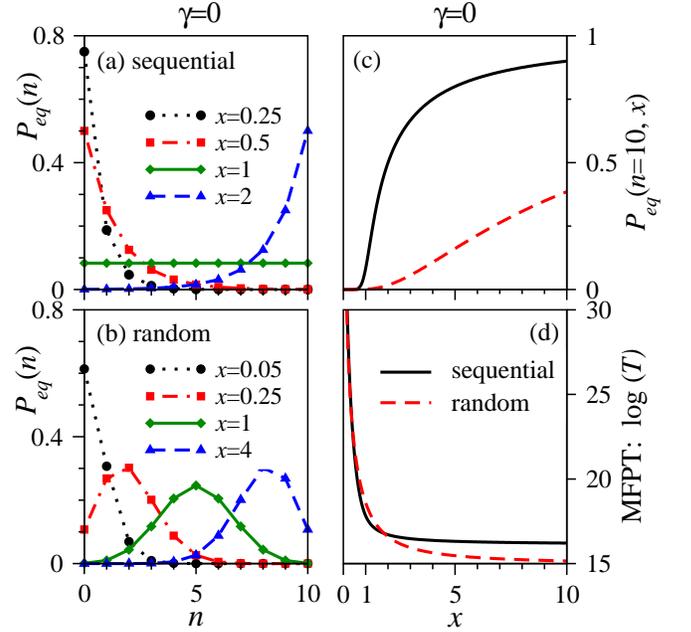}
\end{center}
\vspace{-3mm}
\caption{Equilibrium occupation probabilities $P_{eq}(n)$ for $N=10$
and $\g =0$ as a function of ligand loading $n$ and various affinities
$x$. (a) Sequential binding. (b) Random binding.  Only the random model
exhibits a probability maximum at intermediate $n$. This results from
enhanced opposing drifts at small and large occupancies. The
distributions for $\g \geq 1$ are qualitatively similar to those in
(a). (c) The equilibrium probability of full occupancy ($n=N=10$) for
both models and $\g =0$ as a function of $x$.  Full occupancy is more
likely for the sequential case than for the random one. (d) The
sequential and random case MFPT at $N=10$ and $\g=0$. Note the
logarithmic scale.}
\label{PANEL}
\end{figure}

\begin{figure}[t]
\vspace{2mm}
\begin{center} 
\includegraphics[height=1.55in]{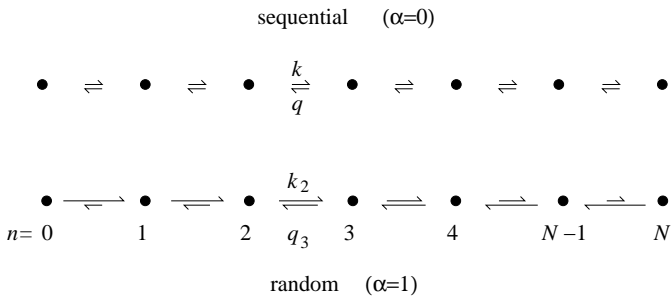}
\end{center}
\caption{Kinetic steps for $x\sim 1$ and $\gamma=0$.  (a) Sequential
process.  All successive steps incrementing the number of bound
receptors have the same rate $k$, while all decrement steps have the
same rates $q$.  (b) Random process. There are more ways to attach
a ligand when $n$ is small. Similarly, there are more ways of removing
receptors when there are many bound receptors.  The effective rates
push the system from both directions toward an intermediate occupation
level.  Given an initial condition at $n=0$, the first passage
statistics to $n=N$ will be controlled more by the effective forward
rates $k_{n}$ than the effective backward rates $q_{n}$. Therefore,
the larger $k_{n}$ tending to load the receptor outweigh the effects
of the larger $q_{n}$ tending for desorption. The overall MFPT is thus
larger for in the random case than in the sequential one.}
\label{LINE}
\end{figure}

\noindent
For the noncooperative case, the occupation probabilities can be
expressed in terms of a binding affinity $x \equiv z e^{\ve_{0}}$.  In
Figs.\,\ref{PANEL}a-b we show the noncooperative $P_{eq}(n)$ for
$N=10$ and various $x$.  Only the random case allows an intermediate
maximum of $P_{eq}(n)$ to develop as $x$ is increased.  From
Eq.\,\ref{RATES}, it can be seen that for small $n$ values the ratio
$k_{n}/q_{n+1}$ is larger for the random case than for the sequential
one.  Conversely, the same ratio is smaller for large $n$.  The random
process thus induces an effective drift towards intermediate $n$
values. This is schematically shown in Fig.\,\ref{LINE}.  We can
rewrite Eq.\,\ref{RATES} in the form $k_{n}/q_{n+1} = \exp(-\Delta
\bar{G}_{n})$, where

\vspace{-2mm}
\begin{equation}
\Delta\bar{G}_{n} = \Delta G_{n}-\a \ln\left({N-n \over n+1}\right)-\ln z,
\end{equation}

\noindent is the effective free energy change that
includes the entropy associated with ligand binding and unbinding.
Unlike $G_n$, the total effective free energy as a function of
$n$,
\begin{equation}
\bar{G}_n = G_n - \a  \ln {N \choose n}-n\ln z,
\end{equation}

\noindent 
can have a minimum with respect to $n$ if $\a = 1$.  Thus, a
maximum in $P_{eq}(n)$ can occur only in the random case.  Equilibrium
distributions and occupancy probabilities have been widely used in
approximations of kinetics of biochemical reactions \cite{VANKAMPEN}.
One commonly used metric is the filling fraction $f_{eq}(x; \alpha,
\gamma) = N^{-1} \sum_{n=1}^{N} n P_{eq}(n)$.  
We can compute $f_{eq}$ under either the sequential or random 
assumptions:

\vspace{-2mm}
\begin{eqnarray}
\label{HILL0}
f_{eq}(x;0,0) &=&
\frac{x (1 - (N+1) x^N + N x^{N+1})}{N (x-1)^2 + N x(x-1) (x^N-1)}  \\
\nonumber \\
f_{eq}(x;1,0) &=&  \frac {x} {1+ x}, 
\label{HILL}
\end{eqnarray}

\noindent respectively. These two expressions coincide only in the infinite
affinity limit $x\rightarrow \infty$, signifying that the process no
longer depends on binding order.  Expressions such as $f_{eq}$ have
been used in kinetic equations describing binding to
hemoglobin/myoglobin \cite{HEMO1,HEMO2}, regulatory networks
\cite{GENE}, viral infection dynamics \cite{VIRAL}, and cell signaling
\cite{SIGNAL}. However, using equilibrium expressions in kinetic
equations assumes separation of time scales or near
steady-states. Moreover, filling fractions do not provide information
about typical loading time scales.  The queuing nonequilibrium
properties are more telling, in this context, than their equilibrium
counterparts and surprising effects may arise.  In particular, faster
complete occupation of a receptor can occur even though the mean
occupancy is lower at equilibrium.
The mean first passage time (MFPT) to a target filling 
can be evaluated by using standard techniques \cite{GARDINER}.
For a receptor in the unoccupied $n=0$ state at $t=0$ 
the MFPT to reach an $n^{*}$ filling is  \cite{PURY}

\vspace{-2mm}
\begin{equation}
T(n^{*};\a) = \sum_{n=0}^{n^*-1}{1 \over k_{n}}
\left(1+\sum_{j=0}^{n-1}\prod_{i=j}^{n-1}{q_{i+1}\over k_{i}}\right).
\label{T0}
\end{equation}

\noindent 
To evaluate Eq.\,(\ref{T0}) we need independent expressions for $k_n$
and $q_n$.  We let $k_n = \nu \, v [L] (N-n)^\a$ and $q_{n+1} = \nu \,
(n+1)^\a \, \exp\left[\Delta G_{n} \right]$, where $\nu$ is the
frequency of reaction attempts.  We choose $k_n$ to be independent of
$\Delta G_n$ because for many systems, such as antibody-antigen 
complexes \cite{KULIN}, cooperative effects are felt only in the detachment
process.  In these cases, the strength of a ligand-receptor chemical
bond depends on the total number of bound ligands present.
These existing ligands typically do not affect the kinetics
of attachment.  For the noncooperative case ($\g=0$), the MFPT to full
occupancy in units of $\nu^{-1}$ is \cite{NOTETOPURY}:
\begin{eqnarray}
\label{fpta}
T(N;\a=0)  = {x \over z(x-1)} \left[\frac {x^{-N}-1} 
{x-1} +N\right]. 
\end{eqnarray}

\begin{figure}[htb]
\vspace{-2mm}
\begin{center}
\includegraphics[height=2.2in]{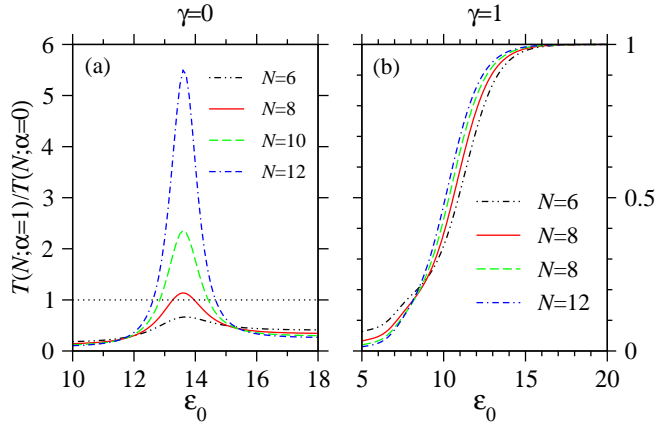}
\end{center}
\vspace{-4mm}
\caption{Mean first passage time ratios
$ T(N;1)/T(N;0)$.
(a) For $\g=0$, random
binding ($\a=1$) is faster except for a window of binding energies
near $x=z e^{\varepsilon_0}=1$. 
For clarity we plot $T(N;1)/T(N;0) $ as a function of $\varepsilon_0$ 
for $z = 10^{-6}$. The critical value $x = 1$ occurs at
$\varepsilon_0 = 13.8$.
The peak scales as $\sim 2^N/N^3$.
(b) For $\g=1$ (and in fact
$\g\geq 1$), the random process always reaches full occupancy faster
than the sequential one.}
\label{RATIOS}
\end{figure}

\noindent
Logarithms of the MFPT from $n=0$ to complete filling $n^* = N $ are
plotted in Fig.\,\ref{PANEL}d for $\g =0$.  A comparison with
Fig.\,\ref{PANEL}c shows an unexpected effect.  Although the
noncooperative random process yields a lower equilibrium ligand
population than the sequential process, its MFPT is shorter.  This is
due to an overall effective increase in the local diffusivity as
evident from the form $k_{n} \propto (N-n)^{\a}$.  However, near the
half-filling affinity $x= 1$, corresponding to $f_{eq}=1/2$, the MFPT
of the random case is greater than that of the sequential one. This is
emphasized in Fig.\,\ref{RATIOS}a, where we plot the ratio of random
to sequential MFPT as a function of binding energy.  A striking peak
develops for $x=1$ provided the receptor has at least $N_{c} = 8$
binding sites.  Indeed, as $N > N_c$ increases, the random MFPT
increases dramatically.
Why does this happen?  From Fig.\,\ref{LINE}, we see that for the
random case near $x=1$, the backward rates $q_{n+1}$ balance the
forward rates $k_{n}$.  This results in a vanishing net drift at
intermediate occupations.  The effective potential $\bar{G}_n$
develops a minimum that for $x=1$ is exactly located at $N/2$.  The
opposing drifts bias the filling to intermediate values forming a
kinetic trap and delaying first saturation. The MFPT increases
significantly due to the time required to escape this trap, despite
the increase in effective diffusivity near the potential minimum.
When $N\geq 8$, the trapping is strong enough that the MFPT of the
random case is greater than that of the sequential process.  This
effect is disrupted as $x$ deviates from unity as the minimum in $\bar
G_n$ disappears.  Note that this behavior occurs only in the {\it
absence} of cooperative effects: for {\it e. g.,} $\g=1$ the
equilibrium occupancies follow the same trends as in
Fig.\,\ref{PANEL}c and the random system reaches full occupancy first,
even at $x = 1$.

\begin{figure}
\includegraphics[height=2.0in]{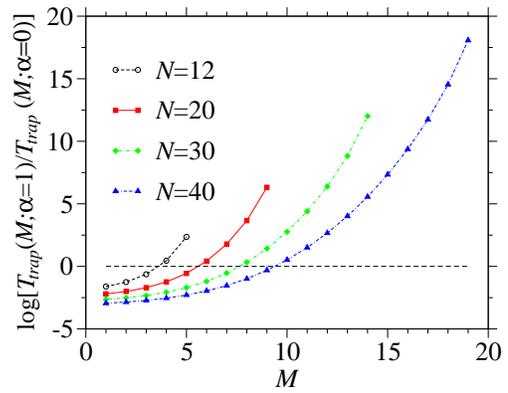}
\caption{Logarithm of the ratio $T_{trap}(M;1)/T_{trap}(M;0)$.
Here, $x=1$, $\g=0$ and the initial occupancy is at $n = N/2$.
Negative values of the ratio imply that escaping a $2M$ kinetic trap
is faster for the random process than for the sequential one. This
occurs only for small enough traps centered on $N/2$ where the random
dynamics is faster. For large enough values of trap length $2M$ the
effects described in Fig.\,\ref{LINE} take place and the exit time of
the random case, $T_{trap}(M;1)$, is much larger than the sequential
one, $T_{trap}(M;0)$.}
\label{TRAP}
\end{figure}

To explicitly see a size-dependent trap, consider the MFPT,
$T_{trap}(M;\a)$, out of a band of sites of length $2M$, centered
about $n=N/2$. For $x=1$, $\g=0$, and an initial condition of $n=N/2$,
the logarithm of the ratio of the random $T_{trap}(M;1)$ to the
sequential $T_{trap}(M;0)$ is plotted in Fig.\,\ref{TRAP}. The
trapping effect is evident from the fact that for large enough $M$ the
random $T_{trap}(M;1)$ can easily be $e^{10}$ times larger than the
sequential $T_{trap}(M;0)$.


Our analysis reveals the conspicuous effects that cooperativity and
sequence have on binding kinetics, especially on the MFPT to
saturation.  The key result is that equilibrium and dynamic
measurements offer different answers as to whether the sequential or
the random order is more efficient in ligand saturation.  At
equilibrium, sequential processes are more likely than random ones to
saturate the ligand.  For dynamic properties, such as MFPTs, random
processes are faster. This is true except for large noncooperative
systems ($N\geq 8$) with a binding affinity near $x=1$, where random
processes dramatically slow down.  One can also consider the reverse
process of emptying a completely filled receptor.  The clearance time
can be obtained by using particle-hole symmetry and the replacements
$k_p^{*} / q_{p+1}^{*} = [(N-p)/(p+1)]^{\alpha} \exp(\Delta G_p) /z$
where $p$ is the number of holes present.

The rate parameters used in our study can also be reinterpreted as the
inverse of the mean servicing time in a customer service queue. Our
results indicate a rich set of outcomes depending on queue size,
cooperativity, and sequentiality.  Levels of ``service discipline,''
or how to order customer service \cite{QUEUE1}, can be implemented in
different ways to achieve the desired outcome with highest
probability, measured by average (equilibrium) or MFPT
(nonequilibrium) attributes.

MRD thanks S. Vinogradov for helpful comments and the ARO for support
through grant DAAD19-02-1-0055. TC acknowledges support from the NSF
and NIH via grants DMS-0206733, DMS-0349195, and K25AI056872.


\begin{references}
\bibitem{NOTETOPURY} Note that Eqs.\,\ref{T0}, \ref{fpta} are derived
in Ref.\,\cite{PURY} as Eqs.\,(2.9), (6.2) respectively, with the
substitutions $M, w_{-M}^{-} \rightarrow 0$, $\gamma \rightarrow
x^{-1}$, and $L \rightarrow N-1$.
\end{references}

\end{document}